\documentclass[12pt]{article}
\usepackage{graphicx}
\begin{document}

\begin{center}

{\Large \bf  Einstein, Wigner, and Feynman: \\[1ex]
  From E = mc$^{2}$ to
Feynman's decoherence  \\[1ex]
via Wigner's little groups}\\ [3ex]

 Y. S. Kim \footnote{electronic mail: yskim@physics.umd.edu}\\[1ex]
{\it Department of Physics, University of Maryland, \\
           College Park, Maryland 20742, U.S.A.}

\end{center}

\vspace{5mm}

\begin{abstract}
The 20th-century physics starts with Einstein and ends with
Feynman.  Einstein introduced the Lorentz-covariant world with
$E = mc^{2}$.  Feynman observed that fast-moving hadrons consist of
partons which act incoherently with external signals.  If quarks
and partons are the same entities observed in different Lorentz frames,
the question then is why partons are incoherent while quarks are coherent.
This is the most puzzling question Feynman left for us to solve.
In this report, we discuss Wigner's role in settling this question.
Einstein's $E = mc^{2}$, which takes the form $E = \sqrt{m^{2} + p^{2}}$,
unifies the energy-momentum relations for massive and massless particles,
but it does not take into account internal space-time structure of
relativistic particles.  It is pointed out Wigner's 1939 paper on the
inhomogeneous Lorentz group defines particle spin and gauge degrees of
freedom in the Lorentz-covariant world.  Within the Wigner framework,
it is shown possible to construct the internal space-time structure for
hadrons in the quark model.  It is then shown that the quark model
and the parton model are two different manifestations of the same
covariant entity.  It is shown therefore that the lack of coherence
in Feynman's parton picture is an effect of the Lorentz covariance.

\end{abstract}

\newpage

\section{Introduction}\label{intro}
Let us start with Einstein. If the momentum of a particle is much
smaller than its mass, the
energy-momentum relation is $E = p^{2}/2m + mc^{2}$.  If
the momentum is much larger than the mass, the relation is $E = cp$.
These two different relations can be combined into one covariant
formula.  This aspect of Einstein's $E = mc^{2}$ is well known.

In the quantum world, particles have internal
space-time variables.  Massive particles have spins while massless
particles have their helicities and gauge variables.  Our first
question is whether this aspect of space-time variables can be
unified into one covariant concept.  The answer to this question is Yes.
Wigner's little group does the job, as is illustrated in Table~\ref{fcmc2}.

In addition, particles can have space-time extensions.  For instance,
in the quark model, hadrons are bound states of quarks.  However, the
hadrons appear as collections of partons when they move with speed
close to the velocity of light.  Quarks and partons seem to have quite
distinct properties.  The most serious difference is that the partons
interact incoherently with external signals while the quark are
coherent particles.   The purpose of this report is to address this
issue, after reviewing what Wigner did and what Feynman did to
understand the Lorentz-covariant world.

By ``further contents of $E = mc^{2}$'', we mean that the
internal space-time structures of massive and massless particles can
be unified into one covariant package, as $E = \sqrt{m^{2} + p^{2}}$
does for the energy-momentum relation.  The mathematical framework of
this program was developed by Eugene Wigner in 1939~\cite{wig39}.
He constructed the maximal subgroups of the Lorentz group whose
transformations will leave the four-momentum of a given particle
invariant.  These groups are known as Wigner's little groups.

Thanks to high-energy accelerators, we can do experiments with massive
particles, such as protons and heavy ions, which move with relativistic
speed. After Gell-Mann invented the quark model where all
hadrons are quantum bound-states of quarks, Feynman came up with an
idea that a hadron appears like a collection of partons when it moves
with a velocity close to that of light.  Then the question is whether
the quark model and the parton model are two different manifestations of
the same covariant entity.

In order to have a theory of extended particles, we need bound-state
wave functions of the quarks inside the hadron.  These wave functions
have to be covariant.  This is the most fundamental problem.  Neither
the present form of quantum mechanics nor the quantum field theory
addresses this issue.  Let us start
with a well-localized wave function in one Lorentz frame.  Then how would
this look to an observer in a different Lorentz frame?  Here, Feynman
was right in guessing that the first covariant wave function has to
be that of harmonic oscillators, as in the case of most of new
theories.  He and his coauthors started constructing such functions,
and showed that the hadronic mass spectra are consistent with the
degeneracies of the three-dimensional oscillators.

However, Feynman {\it et al.} did not succeed in constructing a covariant
formalism.  Indeed, this is possible if we construct Wigner's little
group for massive particles.  The wave functions in this representation
is covariant, and we use these wave functions to show that the quark
model and the parton model are two different manifestations of one
covariant model.  The scope of this report is summarized in
Table~\ref{fcmc2}

\begin{table}

\caption{Further contents of Einstein's $E = mc^{2}$.}\label{fcmc2}
\begin{center}
\begin{tabular}{rccc}
\hline \\[-3.9mm]
\hline
{} & {} & {} & {}\\
{} & Massive, Slow \hspace*{1mm} & COVARIANCE \hspace*{1mm}&
Massless, Fast \\[4mm]\hline
{} & {} & {} & {}\\
Energy- & {}  & Einstein's & {} \\
Momentum & $E = p^{2}/2m$ & $ E = [p^{2} + m^{2}]^{1/2}$ & $E = cp$
\\[4mm]\hline
{} & {} & {} & {}\\
Internal & $S_{3} $ & {}  &  $S_{3}$ \\[-1mm]
space-time & {} & Wigner's  & {} \\ [-1mm]
symmetry & $S_{1}, S_{2}$ & Little Group & Gauge
Trans. \\[4mm]\hline
{} & {} & {} & {}\\
Relativistic & {} & {} & {} \\[-1mm]
Extended & Quark Model & Covariant Model & Partons \\ [-1mm]
Particles & {} & {} & {} \\[2mm]
\hline
\hline
\end{tabular}
\end{center}

\end{table}

\section{Formulation of the Problem}\label{littleg}

It was Eugene Wigner who observed that the space-time symmetry of
relativistic particles is dictated by the Poincar\'e group, the
group of inhomogeneous Lorentz transformations, namely Lorentz
transformations preceded or followed by space-time
translations~\cite{wig39}.  In particular,
Wigner studied the maximal subgroups of the Lorentz group
whose transformations leave the four-momentum of a given free
particle.  These subgroups are called the little groups.
Since the little group leaves the four-momentum invariant, it governs
the internal space-time symmetries of relativistic particles.  Wigner
shows in his paper that the internal space-time symmetries of massive
and massless particles are dictated by the little groups which are
locally isomorphic to the three-dimensional rotation group and the
two-dimensional Euclidean groups respectively.

The group of Lorentz transformations consists of three boosts and
three rotations.  The rotations therefore constitute a subgroup of
the Lorentz group.  If a massive particle is at rest, its four-momentum
is invariant under rotations.  Thus the little group for a massive
particle at rest is the three-dimensional rotation group.  Then what is
affected by the rotation?  The answer to this question is very simple.
The particle in general has its spin.  The spin orientation is going
to be affected by the rotation!

If we use the four-vector coordinate $(x, y, z, t)$, the Lorentz
group is generated by three rotation generators $J_{i}$ and three
boost generators $K_{i}$.  They satisfy the commutation relations
\begin{equation}\label{o3com}
[J_{i}, J_{j}] = i\epsilon_{ijk} J_{k} , \quad
[J_{i}, K_{j}] = i\epsilon_{ijk} K_{k} , \quad
[K_{i}, K_{j}] = -i\epsilon_{ijk} J_{k} .
\end{equation}
This means that the three rotation generators form a closed set of
commutation relations.  Indeed, they are the generators of the
$O(3)$-like little group for a massive particle at rest.  If the
particle is at rest, its momentum is invariant under rotations.
However, its spin direction becomes rotated.  Therefore, the $O(3)$-like
little group defined the spin degree of freedom.

It is not possible to bring a massless particle to its rest frame, but
we can consider a massive particle moving along the $z$ direction
wihout loss of generality.  In his 1939 article, Wigner observed that
the little group for this massless particle is generated by
\begin{equation}\label{n1n2}
J_{3},\qquad, N_{1} = K_{1} - J_{2} , \qquad N_{2} = K_{2} + J_{1} ,
\end{equation}
They satisfy the commutation relations
\begin{equation}\label{e2lcom}
[N_{1}, N_{2}] = 0 , \qquad [J_{3}, N_{1}] = iN_{2} ,
\qquad [J_{3}, N_{2}] = -iN_{1} .
\end{equation}

In order to understand the mathematical basis of the above commutation
relations, let us consider transformations on a two-dimensional plane
with the $xy$ coordinate system.  We can then make rotations around
the origin and translations along the $x$ and $y$ directions.  If we
write these generators as $L, P_{x}$ and $P_{y}$ respectively, they
satisfy the commutation relations~\cite{knp86}
\begin{equation}\label{e2com}
[P_{x}, P_{y}] = 0 , \qquad [L, P_{x}] = iP_{y} ,
\qquad [L, P_{y}] = -iP_{x} .
\end{equation}
This is a closed set of commutation relations for the generators of the
$E(2)$ group.  If we replace $N_{1}$ and $N_{2}$ of Eq.(\ref{e2lcom})
by $P_{x}$ and $P_{y}$, and $J_{3}$ by $L$, the commutations relations
for the generators of the $E(2)$-like little group becomes those for
the $E(2)$-like little group.  This is precisely why we say that
the little group for massless particles are like $E(2)$.

It is not difficult to associate the rotation generator $J_{3}$ with
the helicity degree of freedom of the massless particle.   Then what
physical variable is associated with the $N_{1}$ and $N_{2}$
generators?  Indeed, Wigner was the one who discovered the existence
of these generators, but did not give any physical interpretation to
these translation-like generators.  For this reason, for many years,
only those representations with the zero-eigenvalues of the $N$
operators were thought to be physically meaningful
representations~\cite{wein64}.  It was not until 1971 when Janner
and Janssen reported that the transformations generated by these
operators are gauge transformations~\cite{janner71,kim97poz}.  The
role of this translation-like transformation has also been studied
for spin-1/2 particles, and it was concluded that the polarization
of neutrinos is due to gauge invariance~\cite{hks82,kim97min}.

\section{Contraction of O(3)-like to E(2)-like Little
Groups}\label{contrac}
The $O(3)$-like little group remains $O(3)$-like when the particle is
Lorentz-boosted.  Then, what happens when the particle speed becomes
the speed of light?  The energy-momentum relation
$E = \sqrt{m^{2} + p^{2}}$ become $E = p$.  Is there then a limiting
case of the $O(3)$-like little group?  Since those little groups are
like the three-dimensional rotation group and the two-dimensional
Euclidean group respectively, we are first interested in whether
$E(2)$ can be obtained from $O(3)$.  This will then give a clue to
obtaining the $E(2)$-like little group as a limiting case of
$O(3)$-like little group.  With this point in mind, let us look into
this geometrical problem.

In 1953, Inonu and Wigner formulated this problem as the contraction
of $O(3)$ to $E(2)$~\cite{inonu53}.  Let us see what they did.  We
always associate the three-dimensional rotation group with a spherical
surface.  Let us consider a circular area of radius 1 kilometer centered
on the north pole of the earth.  Since the radius of the earth is more
than 6,450 times longer, the circular region appears flat.  Thus, within
this region, we use the $E(2)$ symmetry group for this region.  The
validity of this approximation depends on the ratio of the two radii.

How about then the little groups which are isomorphic to $O(3)$ and
$E(2)$?  It is reasonable to expect that the $E(2)$-like little group
be obtained as a limiting case for of the $O(3)$-like little group
for massless particles.  In 1981, it was observed by Ferrara and Savoy
that this limiting process is the Lorentz boost~\cite{ferrara82}.
In 1983, using the same limiting process as that of Ferrara and Savoy,
Han {\it et al} showed that transverse rotation generators become the
generators of gauge transformations in the limit of infinite momentum
and/or zero mass~\cite{hks83pl}.

Let us see how this happens when the system is Lorentz-boosted along
the $z$ direction. The $J_{3}$ generator is not affected by
the boost whose transformation matrix takes the form
\begin{equation}
B = \exp{\left(-i\eta K_{3} \right)} .
\end{equation}
On the other hand, the $J_{1}$ and $J_{2}$ matrices become
\begin{equation}
N_{1} = e^{-\eta} B^{-1} J_{2} B ,
\qquad N_{2} = -e^{-\eta} B^{-1} J_{1} B ,
\end{equation}
and they become $N_{1}$ and $N_{2}$ given in Eq.(\ref{n1n2}).  The
generators $N_{1}$ and $N_{2}$ are the contracted $J_{2}$ and $J_{1}$
respectively in the infinite-momentum/zero-mass limit.  In 1987, Kim
and Wigner studied this problem in more detail and showed that the
little group for massless particles is the cylindrical group which is
isomorphic to the $E(2)$ group~\cite{kiwi87jm}.

This completes the second row in Table~\ref{fcmc2}, where Wigner's
little group unifies the internal space-time symmetries of massive and
massless particles.  The transverse components of the rotation generators
become generators of gauge transformations in the
infinite-momentum/zero-mass limit.

\section{Covariant Harmonic Oscillators}\label{covham}

We are now interested in constructing the third row in Table I.  As we
promised in Sec.~\ref{intro}, we will be dealing with hadrons which
are bound states of quarks with space-time extensions.  For this
purpose, we need a set of covariant wave functions consistent with the
existing laws of quantum mechanics, including of course the uncertainty
principle and probability interpretation.  The first wave function which
comes to our mind is the harmonic oscillator wave function.  If we are
interested in Lorentz-transforming them, the most straight-forward
method is to construct representations of the Poincar\'e group using
harmonic oscillators wave functions~\cite{dir45,yuka53,markov56,knp86}.

In this report, we start with the Lorentz-invariant differential
equation of Feynman, Kislinger, and Ravndal~\cite{fkr71}.  It is a
linear partial differential equation which has many different
solutions depending on boundary conditions.  Unlike in the case of
Feynman {\it et al}., we use normalizable wave functions which
constitute a representation of the $O(3)$-like little
group~\cite{knp86}.

Let us consider a bound state of two particles.  For convenience, we
shall call the bound state the hadron, and call its constituents quarks.
Then there is a Bohr-like radius measuring the space-like separation
between the quarks.  There is also a time-like separation between the
quarks, and this variable becomes mixed with the longitudinal spatial
separation as the hadron moves with a relativistic speed.  There are
no quantum excitations along the time-like direction.  On the other
hand, there is the time-energy uncertainty relation which allows
quantum transitions.  It is possible to accommodate these aspect within
the framework of the present form of quantum mechanics.  The uncertainty
relation between the time and energy variables is the c-number
relation~\cite{dir27}, which does not allow excitations along the
time-like coordinate.  We shall see that the covariant harmonic
oscillator formalism accommodates this narrow window in the present
form of quantum mechanics.

For a hadron consisting of two quarks, we can consider their space-time
positions $x_{a}$ and $x_{b}$, and use the variables
\begin{equation}
X = (x_{a} + x_{b})/2 , \qquad x = (x_{a} - x_{b})/2\sqrt{2} .
\end{equation}
The four-vector $X$ specifies where the hadron is located in space and
time, while the variable $x$ measures the space-time separation between
the quarks.  In the convention of Feynman {\it et al.}~\cite{fkr71},
the internal motion of the quarks bound by a harmonic oscillator
potential of unit strength can be described by the Lorentz-invariant
equation
\begin{equation}\label{osceq}
{1\over 2}\left\{x^{2}_{\mu} -
{\partial ^{2} \over \partial x_{\mu }^{2}}
\right\} \psi (x)= \lambda \psi (x) .
\end{equation}
It is now possible to construct a representation of the Poincar\'e group
from the solutions of the above differential equation~\cite{knp86}.

The coordinate $X$ is associated with the overall hadronic
four-momentum, and the space-time separation variable $x$ dictates
the internal space-time symmetry or the $O(3)$-like little group.  Thus,
we should construct the representation of the little group from the
solutions of the differential equation in Eq.(\ref{osceq}).  If the
hadron is at rest, we can separate the $t$ variable from the equation.
For this variable we can assign the ground-state wave function to
accommodate the c-number time-energy uncertainty relation~\cite{dir27}.
For the three space-like variables, we can solve the oscillator
equation in the spherical coordinate system with usual orbital and
radial excitations.  This will indeed constitute a representation of
the $O(3)$-like little group for each value of the mass.  The solution
should take the form
\begin{equation}
\psi (x,y,z,t) = \psi (x,y,z) \left({1\over \pi }\right)^{1/4}
\exp \left(-t^{2}/2 \right) ,
\end{equation}
where $\psi(x,y,z)$ is the wave function for the three-dimensional
oscillator with appropriate angular momentum quantum numbers.  Indeed,
the above wave function constitutes a representation of Wigner's
$O(3)$-like little group for a massive particle~\cite{knp86}.

Since the three-dimensional oscillator differential equation is
separable in both spherical and Cartesian coordinate systems,
$\psi(x,y,z)$ consists of Hermite polynomials of $x, y$, and $z$.
If the Lorentz boost is made along the $z$ direction, the $x$ and $y$
coordinates are not affected, and can be temporarily dropped from the wave
function.  The wave function of interest can be written as
\begin{equation}
\psi^{n}(z,t) = \left({1\over \pi }\right)^{1/4}\exp \pmatrix{-t^{2}/2}
\psi_{n}(z) ,
\end{equation}
with
\begin{equation}
\psi ^{n}(z) = \left({1 \over \pi n!2^{n}} \right)^{1/2} H_{n}(z)
\exp (-z^{2}/2) ,
\end{equation}
where $\psi ^{n}(z)$ is for the $n$-th excited oscillator state.
The full wave function $\psi ^{n}(z,t)$ is
\begin{equation}\label{2.6}
\psi ^{n}_{0}(z,t) = \left({1\over \pi n! 2^{n}}\right)^{1/2} H_{n}(z)
\exp \left\{-{1\over 2}\left(z^{2} + t^{2} \right) \right\} .
\end{equation}
The subscript $0$ means that the wave function is for the hadron at
rest.  The above expression is not Lorentz-invariant, and its
localization undergoes a Lorentz squeeze as the hadron moves along the
$z$ direction~\cite{knp86}.

It is convenient to use the light-cone variables to describe Lorentz
boosts.  The light-cone coordinate variables are
\begin{equation}
u = (z + t)/\sqrt{2} , \qquad v = (z - t)/\sqrt{2} .
\end{equation}
In terms of these variables, the Lorentz boost along the $z$
direction,
\begin{equation}
\pmatrix{z' \cr t'} = \pmatrix{\cosh \eta & \sinh \eta \cr
\sinh \eta & \cosh \eta } \pmatrix{z \cr t} ,
\end{equation}
takes the simple form
\begin{equation}\label{lorensq}
u' = e^{\eta } u , \qquad v' = e^{-\eta } v ,
\end{equation}
where $\eta $ is the boost parameter and is $\tanh ^{-1}(v/c)$.
Indeed, the $u$ variable becomes expanded while the $v$ variable becomes
contracted.  This is the squeeze mechanism illustrated discussed
extensively in the literature~\cite{kn73,knp91}.

The wave function of Eq.(\ref{2.6}) can be written as
\begin{equation}\label{10}
\psi ^{n}_{o}(z,t) = \psi ^{n}_{0}(z,t)
= \left({1 \over \pi n!2^{n}} \right)^{1/2} H_{n}\left((u + v)/\sqrt{2}
\right) \exp \left\{-{1\over 2} (u^{2} + v^{2}) \right\} .
\end{equation}
If the system is boosted, the wave function becomes
\begin{equation}\label{11}
\psi ^{n}_{\eta }(z,t) = \left({1 \over \pi n!2^{n}} \right)^{1/2}
H_{n} \left((e^{-\eta }u + e^{\eta }v)/\sqrt{2} \right)
\times \exp \left\{-{1\over 2}\left(e^{-2\eta }u^{2} +
e^{2\eta }v^{2}\right)\right\} .
\end{equation}

In both Eqs. (\ref{10}) and (\ref{11}), the localization property of
the wave function in the $u v$ plane is determined by the Gaussian factor,
and it is sufficient to study the ground state only for the essential
feature of the boundary condition.  The wave functions in Eq.(\ref{10})
and Eq.(\ref{11}) then respectively become
\begin{equation}\label{13}
\psi _{0}(z,t) = \left({1 \over \pi} \right)^{1/2}
\exp \left\{-{1\over 2} (u^{2} + v^{2}) \right\} .
\end{equation}
If the system is boosted, the wave function becomes
\begin{equation}\label{14}
\psi _{\eta }(z,t) = \left({1 \over \pi }\right)^{1/2}
\exp \left\{-{1\over 2}\left(e^{-2\eta }u^{2} +
e^{2\eta }v^{2}\right)\right\} .
\end{equation}
We note here that the transition from Eq.(\ref{13}) to Eq.(\ref{14})
is a squeeze transformation.  The wave function of Eq.(\ref{13}) is
distributed within a circular region in the $u v$ plane, and thus in
the $z t$ plane.  On the other hand, the wave function of Eq.(\ref{14})
is distributed in an elliptic region.  This is how the wave function is
Lorentz-boosted.

\section{Feynman's Parton Picture}\label{parton}

It is safe to believe that hadrons are quantum bound states of quarks having
localized probability distribution.  As in all bound-state cases, this
localization condition is responsible for the existence of discrete mass
spectra.  The most convincing evidence for this bound-state picture is the
hadronic mass spectra which are observed in high-energy
laboratories~\cite{knp86,fkr71}.  However, this picture of bound states
is applicable only to observers in the Lorentz frame in which the hadron
is at rest.  How would the hadrons appear
to observers in other Lorentz frames?

In 1969, Feynman observed that a fast-moving hadron can be regarded as a
collection of many ``partons'' whose properties do not appear to be
identical to those of quarks~\cite{fey69}.  For example, the number of
quarks inside a static proton is three, while the number of partons in a
rapidly moving proton appears to be infinite.  The question then is how
the proton looking like a bound state of quarks to one observer can appear
different to an observer in a different Lorentz frame?  Feynman made the
following systematic observations.

\begin{itemize}

\item[ a).] The picture is valid only for hadrons moving with velocity
     close to that of light.

\item[ b).] The interaction time between the quarks becomes dilated,
  and partons behave as free independent particles.

\item[ c).] The momentum distribution of partons becomes widespread as
  the hadron moves very fast.

\item[ d).] The number of partons seems to be infinite or much larger
  than that of quarks.

\end{itemize}

\noindent Because the hadron is believed to be a bound state of two or
three quarks, each of the above phenomena appears as a paradox,
particularly b) and c) together.  We would like to resolve this paradox
using the covariant harmonic oscillator formalism.

For this purpose, we need a momentum-energy wave function.  If the quarks
have the four-momenta $p_{a}$ and $p_{b}$, we can construct two independent
four-momentum variables~\cite{fkr71}
\begin{equation}
P = p_{a} + p_{b} , \qquad q = \sqrt{2}(p_{a} - p_{b}) .
\end{equation}
The four-momentum $P$ is the total four-momentum and is thus the hadronic
four-momentum.  $q$ measures the four-momentum separation between the quarks.

\begin{figure}[thbp]
\centerline{\includegraphics[scale=0.5]{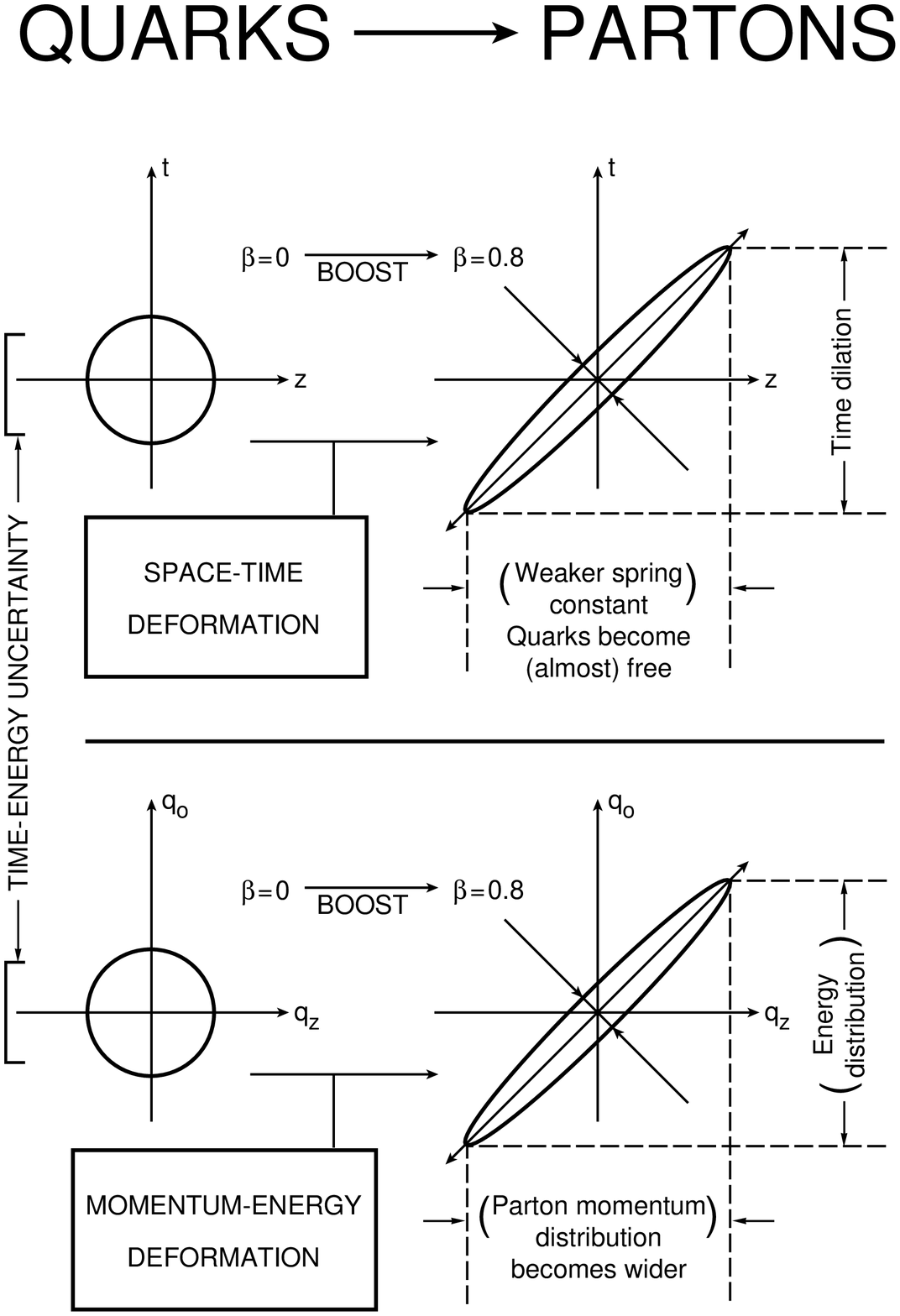}}
\caption{Lorentz-squeezed space-time and momentum-energy wave functions.
As the hadron's speed approaches that of light, both wave functions
become concentrated along their respective positive light-cone axes.
These light-cone concentrations lead to Feynman's parton
picture.}\label{f.parton}
\end{figure}

We expect to get the momentum-energy wave function by taking the Fourier
transformation of Eq.(\ref{14}):
\begin{equation}\label{fourier}
\phi_{\eta }(q_{z},q_{0}) = \left({1 \over 2\pi }\right)
\int \psi_{\eta}(z, t) \exp{\left\{-i(q_{z}z - q_{0}t)\right\}} dx dt .
\end{equation}
Let us now define the momentum-energy variables in the light-cone coordinate
system as
\begin{equation}\label{conju}
q_{u} = (q_{0} - q_{z})/\sqrt{2} ,  \qquad
q_{v} = (q_{0} + q_{z})/\sqrt{2} .
\end{equation}
In terms of these variables, the Fourier transformation of
Eq.(\ref{fourier}) can be written as
\begin{equation}\label{fourier2}
\phi_{\eta }(q_{z},q_{0}) = \left({1 \over 2\pi }\right)
\int \psi_{\eta}(z, t) \exp{\left\{-i(q_{u} u + q_{v} v)\right\}} du dv .
\end{equation}
The resulting momentum-energy wave function is
\begin{equation}\label{phi}
\phi_{\eta }(q_{z},q_{0}) = \left({1 \over \pi }\right)^{1/2}
\exp\left\{-{1\over 2}\left(e^{-2\eta}q_{u}^{2} +
e^{2\eta}q_{v}^{2}\right)\right\} .
\end{equation}
Since we are using the harmonic oscillator, the mathematical form
of the above momentum-energy wave function is identical to that of the
space-time wave function.  The Lorentz squeeze properties of these wave
functions are also the same, as are indicated in Fig.~\ref{f.parton}.
These squeeze transformations perfectly consistent with the algorithms
of the Poincar\'e group~\cite{kim89}.

When the hadron is at rest with $\eta = 0$, both wave functions behave
like those for the static bound state of quarks.  As $\eta$ increases,
the wave functions become continuously squeezed until they become
concentrated along their respective positive light-cone axes.  Let us
look at the z-axis projection of the space-time wave function.  Indeed,
the width of the quark distribution increases as the hadronic speed
approaches that of the speed of light.  The position of each quark
appears widespread to the observer in the laboratory frame, and the
quarks appear like free particles.


Furthermore, interaction time of the quarks among themselves become
dilated.  Because the wave function becomes wide-spread, the distance
between one end of the harmonic oscillator well and the other end
increases as is indicated in Fig.~\ref{f.parton}.  This effect,
first noted by Feynman~\cite{fey69}, is universally observed in
high-energy hadronic experiments.  The period is oscillation is
increases like $e^{\eta}$.  On the other hand, the interaction time
with the external signal, since it is moving in the direction opposite
to the direction of the hadron, it travels along the negative
light-cone axis.  If the hadron contracts along the negative
light-cone axis, the interaction time decreases by $e^{-\eta}$.
The ratio of the interaction time to the oscillator period becomes
$e^{-2\eta}$.  The energy of each proton coming out of the Fermilab
accelerator is $900 GeV$.  This leads the ratio to $10^{-6}$.  This
is indeed a small number.  The external signal is not able to sense
the interaction of the quarks among themselves inside the hadron.
This is the reason why the partons appear to be incoherent to external
signals.  Indeed, Feynman's decoherence is an effect of the Lorentz
covariance.

\section*{Concluding Remarks}
Due to Einstein, this world, at least the physics world, became
Lorentz-covariant.  The lack of coherence in Feynman's parton picture
is the most puzzling question in covariance.  It is a pleasure to report
that Wigner's formulation of the internal space-time symmetries of
relativistic particles provide a resolution to this problem.

In this report, we discussed Wigner's 1939 paper on the representations
of the Poincar\'e group.  Wigner wrote many other papers.  They were
also discussed at this conference.  We are grateful to Professors
Joszef Janszky and Peter Adam for organizing this historical conference.
The author would like to thank Jiri Kvita for pointing out an typographical
error in the original version.

\end{document}